\documentclass[12pt,letterpaper]{article}
\usepackage{graphicx,cite,amsfonts}
\usepackage{fancyhdr}
\usepackage[singlespacing]{setspace}
\usepackage[left=2.5cm,right=2.5cm,bottom=2.5cm,includeheadfoot,letterpaper]{geometry}

\def\rv{{\bf r}}
\def\bv{{\bf b}}

\def\kv{{\bf k}}

\def\(#1){(\ref{#1})}           
\newcommand{\Rset}{\mathbb{R}}
\newcommand{\Zset}{\mathbb{Z}}

\begin{document}
\thispagestyle{plain}
\begin{center}
\Huge{\bf Nanotechnology and Quasicrystals: From self assembly to
  photonic applications.}  

\bigskip
\large{Ron Lifshitz}

\medskip \normalsize{\it Raymond and Beverly Sackler School of Physics
  \& Astronomy\\ Tel Aviv University, 69978 Tel Aviv, Israel.} 
\end{center}

\begin{quote}
  After providing a concise overview on quasicrystals and their
  discovery more than a quarter of a century ago, I consider the
  unexpected interplay between nanotechnology and quasiperiodic
  crystals. Of particular relevance are efforts to fabricate
  artificial functional micro- or nanostructures, as well as efforts
  to control the self-assembly of nanostructures, where current
  knowledge about the possibility of having long-range order without
  periodicity can provide significant advantages. I discuss examples of
  systems ranging from artificial metamaterials for photonic
  applications, through self-assembled soft matter, to surface waves
  and optically-induced nonlinear photonic quasicrystals.
\end{quote}

\section{Nanotechnology and quasicrystals?}

When organizers of the NATO Advanced Research Workshop on
nanotechnology, held in St.~Petersburg in June 2008, asked me to
deliver a keynote lecture on quasicrystals I was certain that they had
made a mistake. I have been studying quasicrystals for over 15 years
and investigating nanomechanical systems for just about a decade, and
although one always finds connections between different scientific
fields, I had never expected such an invitation. Nevertheless, the
organizers insisted and explained that they wanted to learn about the
possibility of exploiting nontrivial symmetries---perhaps imitating
what viruses do---and in general, learn the lesson of a scientific
community that was forced by nature to keep an open mind and ``think
outside of the box''.

This chapter is motivated by my presentation on quasicrystals at the
NATO ARW on nanotechnology in St.~Petersburg. I begin in
Sec.~\ref{discovery} by describing the discovery of quasicrystals and
the scientific revolution that followed. I argue that now that the
surprise has long subsided, and we are well aware of the possibility
for having long-range order without periodicity, we are equipped with
much knowledge that can be adopted in other fields---possibly
nanotechnology. Particularly relevant are efforts to fabricate
artificial functional micro- or nanostructures, as well as efforts to
control the self-assembly of nanostructures. I give an elementary
introduction to quasicrystals in Sec.~\ref{QCs}. I then continue in
Sec.~\ref{meta} with an overview on the use of quasicrystalline
structures in artificial metamaterials for photonic applications, and
proceed in Sec.~\ref{soft} to give a description of our current studies
of the self-assembly of soft matter, namely micelle-forming dendrimers
and polymeric stars, into quasicrystals. I make an interlude in
Sec.~\ref{LP} to an even softer system---parametrically-excited
surface waves in fluids---that spontaneously forms quasicrystalline
patterns. I finish in Sec.~\ref{dynamics} by mentioning some dynamical
properties of quasicrystals that might be more easily investigated in
soft-matter quasicrystals than in solid-state quasicrystals,
introducing yet another physical system, that of optically-induced
nonlinear photonic quasicrystals, which we have been using to study
quasicrystals.

\bigskip
\section{The discovery of quasicrystals as a modern-day\\ scientific
  revolution} 
\label{discovery}

For almost two millennia crystallography was predominantly concerned
with the external morphology of crystals. Crystallographers studied
the naturally-occurring facets of crystals, which always intersect at
precise and characteristic angles. It was only in the 17$^{th}$
century that modern crystallography was born, thanks to the brilliant
idea---attributed to such great scientists as Kepler and Hooke---that
crystal shapes were the result of internal {\it order\/} of ``atomic''
units. In his study of tilings of the plane by polygons,
Kepler~\cite{kepler} was quick to realize that very few regular
polygons---namely the triangle, the square, and the hexagon---can tile
the plane without introducing overlaps or leaving holes. Yet, this
observation did not deter him from constructing a well-ordered tiling
of the plane, consisting of pentagons and decagons that requires some
of the decagons to overlap, and leaves holes in the form of 5-fold
stars---a tiling that was to be rediscovered by Penrose more than 350
years later~\cite{penrose,gardner}. Without realizing it, Kepler had
discovered some of the basic properties of aperiodic order, commenting
in his own words that {\it ``The structure is very elaborate and
  intricate.''}

Unfortunately, by the end of the 18$^{th}$ century, when Ha\"uy began
formulating the mathematical theory of crystallography~\cite{hauy},
Kepler's insightful drawings of aperiodic tilings with decagonal
symmetry were long forgotten. Consequently, mathematical
crystallography was founded upon the premise that the internal order
of crystals was necessarily achieved through a periodic filling of
space.  Thus, crystallography treated {\it order\/} and {\it
  periodicity\/} synonymously, using either property interchangeably
to define the notion of a {\it crystal.}  The periodic nature of
crystals was ``confirmed'' with the discovery of x-ray crystallography
and numerous other experimental techniques throughout the 20$^{th}$
century. As they are more ``elaborate and intricate'' and less
commonly found in Nature, aperiodic crystals were completely
overlooked.  Periodicity became the underlying paradigm, not only for
crystallography itself, but also for other disciplines such as
materials science, solid state chemistry, and condensed matter
physics, whose most basic experimental and theoretical tools rely on
its existence.

This historical oversight was corrected with Shechtman's discovery of
quasicrystals in 1982---a discovery that sparked a {\it bona fide\/}
Kuhnian scientific revolution~\cite{kuhn}, as described by
Cahn~\cite{cahn}, one of the co-authors of the announcement of the
discovery, which appeared only two years later in
1984~\cite{shechtman}.  In the famous Hargittai
interviews~\cite{hargittai} with all the scientists involved in the
initial study of quasicrystals, Mackay is quoted as saying that
\begin{quote}{\it``It's a discovery of a material which breaks the
    laws that were artificially constructed. They were not laws of
    nature; they were laws of the human classificatory
    system.''}\end{quote} Nature had found a way of achieving order
without periodicity, and Shechtman was the first to pay attention to
it, and not to dismiss it as an experimental artifact, as many must
have done before him. He confronted a skeptical scientific community
that was unwilling to relinquish its most basic paradigm that order
stems from periodicity. His biggest challenger was Pauling, one of the
greatest chemists of the 20$^{th}$ century and a leading
crystallographer of that time. In a remarkable article, suggesting an
alternative description of Shechtman's icosahedral quasicrystal as
multiple twinning of periodic cubic crystals---a description that not
much later was shown to be
incorrect~\cite{bancel}---Pauling~\cite{pauling} concluded by saying
that\begin{quote}{\it ``Crystallographers can now cease to worry that
    the validity of one of the accepted bases of their science has
    been questioned''}.\end{quote}

Today, thousands of diffraction diagrams later, compounded by
high-quality experimental data---such as images from high-resolution
transmission electron microscopes and atomic-resolution scanning
tunneling microscopes---the existence of order without periodicity has
been unequivocally established. Not only has the periodicity paradigm
been questioned, as Pauling worried, it has been completely shattered.
By 1992, only a decade after the discovery, the International Union of
Crystallography, through its Commission on Aperiodic
Crystals~\cite{iucr}, was ready to publish a provisional definition of
the term {\it crystal\/} that abolishes {\it periodicity,} and implies
that {\it order\/} should be its replacement. The commission was not
ready to give precise microscopic descriptions of all the ways in
which order can be achieved. Clearly, periodicity is one way of
achieving order, quasiperiodicity as in the Penrose-Kepler tiling is
another, but the committee was uncertain whether there were other ways
that were yet to be discovered.  The Commission opted to shift the
definition from a microscopic description of the crystal to a property
of the data collected in a diffraction experiment. It decided on a
temporary working-definition whereby a {\it crystal\/} is
\begin{quote}{\it ``any solid having an essentially discrete
    diffraction diagram.''}\end{quote} Thus, crystals that are
periodic are now explicitly called {\it periodic crystals,} all others
are called {\it aperiodic crystals.} 

The 1992 definition is consistent with the notion of long-range
order---one of the basic notions of condensed-matter
physics~\cite{anderson,sethna}---dating back to ideas of Landau in
which the symmetry-breaking transition from a disordered
(high-symmetry) phase to an ordered (low-symmetry) phase is quantified
by the appearance of a non-zero {\it order parameter}---in this case,
the appearance of Bragg peaks in the Fourier spectrum. Stated in plain
words, long-range order---or in the context of our current discussion,
long-range positional order---is a measure of the correlations between
the positions of atoms in distant regions of the material. 

The 1992 definition was left sufficiently vague so as not to impose
unnecessary constraints on any further study of crystallinity. Indeed,
the discovery inspired a renaissance in crystallography, and has made
a great impact on mathematics~\cite{senechal-qcsj}, as can be seen by
an ever increasing number of books~\cite{senechal,moody,patera,%
  baakemoody}. Much effort has been invested in studying the
characteristics of order, as well as in the development of diffraction
theory~\cite{hof,lagarias,lenz}, once it was realized that periodicity
was not a necessary condition for order and for the appearance of
Bragg peaks in a diffraction spectrum. In particular, Baake and
several co-workers~\cite{baakeentropy,baakevisible,baakerandom,%
  baakediffuse, baakechapter} have performed a systematic study whose
purpose is to characterize which distributions of matter diffract to
produce a pure point component in their spectrum, and thus can qualify
as possessing long-range order. Sufficient progress has already been
made, that we are now ready to complete the paradigm shift and adopt a
permanent definition of crystal that is firmly based on the notion of
{\it order}~\cite{what}.

In October 2007 the quasicrystal community---consisting of
mathematicians, physicists, chemists, materials scientists, surface
scientists, and even photonics engineers---celebrated the 25$^{th}$
anniversary of the discovery in a ``Silver Jubilee''
conference~\cite{qcsj}.  Today, the science of quasicrystals, with its
growing number of
textbooks~\cite{senechal,moody,patera,baakemoody,steinhardt,janot,%
  stadnik,suck,dubois}, is a mature science. Old paradigms have been
carefully transformed into new ones~\cite{rebirth}; definitions have
been changed~\cite{quasidef}; space-group theory has been generalized
to quasicrystals using two alternative
approaches~\cite{mermin,rabson,physa,volc}, and even extended to treat
novel long-range order possessing color~\cite{rmpcolor} or magnetic
symmetry~\cite{prlmag,led}); and many fundamental problems---including
Bak's famous question: ``Where are the atoms?''~\cite{bakatoms}---are
finding their solutions~\cite{takakura,patNV}.  Nevertheless, other
important questions have remained unanswered to this day. Many of
these---such as the electronic and other physical properties of
quasicrystals~\cite{mayou-discussion}, the surface science of
quasicrystals~\cite{thiel-discussion}, and the importance of the
phason degree of freedom~\cite{widom-discussion}---were hotly debated
at the ``Silver Jubilee'' conference~\cite{qcsj}, and continue to
drive us forward.  One particularly interesting set of questions, and
the focus of this chapter, deals with metamaterials and soft-matter
quasicrystals---the newly added members of the quasicrystal family.

We know today that quasicrystals are more common than one had
originally expected. Scores, or even hundreds, of binary and ternary
{\it metallic alloys\/} are known to form quasicrystalline
phases~\cite{tsai}---mostly with icosahedral or decagonal point-group
symmetry---and more are continuously being discovered.  Nevertheless,
it is only in the last few years that quasicrystals have been
discovered (independently) in two different {\it soft-matter\/}
systems: micelle-forming dendrimers~\cite{zeng1,zeng2,mehl} and
three-armed star block copolymers~\cite{takano1,matsushita,takano2}.
These newly discovered {\it soft quasicrystals\/} not only provide
exciting alternative experimental platforms for the basic study of
quasiperiodic long-range order, but also hold the promise for new
applications based on self-assembled nanomaterials~\cite{scaffolds,%
  electronics,selfassemblyreview}, with unique electronic or photonic
properties that take advantage of the quasiperiodicity, which is
relevant to our focus here.


The current emphasis in the study of soft-matter quasicrystals is to
find an explanation for their thermodynamic stability, and thus learn
how to control their self-assembly. To this date, soft quasicrystals
have been observed only with dodecagonal point-group symmetry. Their
source of stability is therefore likely to be different from their
solid-state siblings, yet a good understanding of the stability of one
quasiperiodic system may help to understand the stability of the
other. In what follows I shall review our initial understanding of
what might be the source of stability of soft quasicrystals, while
providing a concise background on the subject. I will try to emphasize
the important relations between the variety of different physical and
chemical systems that form quasicrystalline phases---atomic
quasicrystals, soft quasicrystals, surface waves, and also
artificially-produced structures and metamaterials.

\section{Quasicrystals---Terminology and general framework}
\label{QCs}

Let us consider a scalar function $\rho(\rv)$ that describes the
electronic density or the ionic potential of a material.  The Fourier
transform of a quasiperiodic density $\rho(\rv)$ has the form
\begin{equation}
  \label{eq:fourier}
  \rho(\rv) = \sum_{\kv\in L} \rho(\kv) e^{i\kv\cdot\rv},
\end{equation}
where the (reciprocal) lattice $L$ is a finitely generated
$\Zset$-module, {\it i.e.}  it can be expressed as the set of all
integral linear combinations of a finite number $D$ of $d$-dimensional
wave vectors, $\bv^{(1)},\ldots,\bv^{(D)}$. In the special case where
$D$, called the rank of the crystal, is equal to the physical
dimension $d$, the crystal is periodic. We refer to all quasiperiodic
crystals that are not periodic as ``quasicrystals''.\footnote[1]{Some
  older texts require crystals to possess so-called ``forbidden
  symmetries'' in order to be regarded as quasicrystals. It is now
  understood that such a requirement is inappropriate.  See
  Ref.~\cite{quasidef} for details, and Ref.~\cite{fibonacci} for
  examples of square and cubic quasicrystals.} This term was first
coined by Levine and Steinhardt~\cite{levine1,levine2} in the first of
a series of important theoretical papers that were published in the
eighties.

It is useful to introduce a physical setting based on the notion of
symmetry breaking that was mentioned
earlier~\cite{anderson,martin,sethna}. Let us assume that the
quasiperiodically-ordered state, described by $\rho(\rv)$, is a
symmetry-broken stable ground state of some generic free energy $\cal
F$, invariant under all translations and rotations in $\Rset^d$.  This
is the same as saying that the physical interactions giving rise to
the quasicrystal are themselves translationally and rotationally
invariant, and that the ground state breaks this symmetry.  The free
energy $\cal F$ is a functional of $\rho(\rv)$, which in Fourier space
takes the general form
\begin{equation}
  \label{eq:free}
  {\cal F}\{\rho\} = \sum_{n=2}^\infty \sum_{\kv_1\ldots\kv_n}
  A(\kv_1,\ldots\kv_n) \rho(\kv_1)\cdots\rho(\kv_n).
\end{equation}
Based on the idea of such a generic free energy, Rokhsar, Wright, and
Mermin~\cite{mermin} introduced the notion of {\it
  indistinguishability,} namely that two functions $\rho(\rv)$ and
$\rho'(\rv)$ are indistinguishable if a generic free energy cannot
distinguish between them and assigns them both the same value.  It
then follows~\cite{mermin,rabson,physa} that $\rho(\rv)$ and
$\rho'(\rv)$ are indistinguishable if and only if
\begin{equation}
  \label{eq:gauge}
  \forall \kv\in L:\quad \rho'(\kv) = e^{2\pi i\chi(\kv)}\rho(\kv),
\end{equation}
where $\chi(\kv)$, called a {\it gauge function,} has the property
that $\chi(\kv_1+\kv_2) \equiv \chi(\kv_1) + \chi(\kv_2)$ whenever
$\kv_1$ and $\kv_2$ are in $L$, where `$\equiv$' denotes equality to
within an additive integer.

Gauge functions are useful in describing the relations between the
different symmetry-broken ground states of $\cal F$. Dr\"ager and
Mermin~\cite{jorg} showed that gauge functions form a vector space
$V^*$ of all real-valued linear functions on the lattice $L$, and
because $L$ has rank $D$, $V^*$ is a $D$-dimensional vector space over
the real numbers. The space $V^*$ contains, as a subspace, all the
integral-valued linear functions on $L$. This subset, which has the
algebraic structure of a rank-$D$ $\Zset$-module (just like $L$
itself) is denoted by $L^*$. Gauge functions in $L^*$ leave the
ground-state density invariant. Gauge functions that belong to the
quotient space $V^*/L^*$ take the ground state described by $\rho$
into a different, yet indistinguishable, ground state described by
some other density function $\rho'$. Thus, one can parameterize all
the related symmetry-broken ground states of $\cal F$ on a simple
$D$-torus---the order parameter space $V^*/L^*$.

Different, yet indistinguishable, ground states may also be related by
rotations $g\in O(d)$. In this case $\rho'$ in (\ref{eq:gauge}) is
simply a rotated version of $\rho$, and for each such rotation $g$
there is a special gauge function $\phi_g$, called a {\it phase
  function,} satisfying
\begin{equation}
  \label{eq:point}
  \forall \kv\in L:\quad \rho(g\kv) = e^{2\pi i\phi_g(\kv)}\rho(\kv).
\end{equation}
The set of all rotations satisfying (\ref{eq:point}) forms the point
group of the crystal, and along with the corresponding phase functions
completely characterizes its space group~\cite{rabson,mermin,physa,%
  rmpcolor,prlmag,led}. Unlike periodic crystals, quasicrystals are
not restricted in the order of their rotational symmetry. The
point-group condition (\ref{eq:point}) is applicable to operations of
any order (as long as the rank of the crystal is finite). Thus, in
general, $g$ may be an $n$-fold rotation of any order $n$.

\section{Exploiting quasicrystalline order in artificially constructed
  metamaterials}
\label{meta}

Interesting applications are starting to emerge lately that take
advantage of quasiperiodic long-range order in metamaterials, or
artificially constructed quasicrystals~\cite{deloudi}. Two main
features distinguish quasicrystals from periodic crystals in the
practical sense of using them as metamaterials. The first and more
obvious is the relaxation of any symmetry constraints. In dealing with
quasicrystals for over a quarter of a century we have learned how to
design structures with axes of symmetry of arbitrarily high order.  Of
course, as the order of symmetry increases, so does the rank of the
crystal and therefore its complexity. Nevertheless, simple rank-4
2-dimensional structures already allow one to construct structures
with axes of 5-fold, 8-fold, 10-fold, and 12-fold symmetry---a
substantial improvement over the limited 2-fold, 3-fold, 4-fold, and
6-fold axes possible with periodic crystals. Most applications of
quasicrysalline metamaterials to date are thus based on this notion.
These are mostly {\it linear photonic crystals,} where quasiperiodic
modulations of the index of refraction of different materials are used
in order to engineer their optical response. These applications take
advantage of the fact that there are no restrictions on the order of
the rotational symmetry in order to obtain nearly-isotropic photonic
band gaps~\cite{jin,zoorob}. Dodecagonal (12-fold) quasicrystals are
particularly useful as they are at the same time quite simple (the
rank is only 4) yet the dodecagon is a far better approximation of a
circle than the hexagon, which is the best one can achieve with
periodic photonic crystals. Initial work is also carried out with {\it
  phononic quasicrystals} for controlling the propagation of sound
waves~\cite{deloudi}.

The second feature of quasicrystals, useful for metamaterial
applications, is the complete relaxation of any constraints on the
positions of Bragg peaks in their diffraction diagrams.  One may
design quasiperiodic metamaterials in which the Bragg peaks are placed
at predetermined positions in Fourier space.  We have exploited this
idea in the {\it nonlinear\/} optical domain~\cite{Lifshitz_PRL2005,
  alon1,alon2}, where recent technological progress has enabled to
modulate the second-order nonlinear susceptibility with micron-scale
resolution in various materials, such as ferroelectrics,
semiconductors, and polymers. In these {\it nonlinear photonic
  crystals\/} the modulation can be achieved by planar techniques,
thereby offering either one or two dimensions for modulation.
Moreover, there are no photonic bandgaps in these metamaterials,
because the first-order susceptibility, and hence the refractive
index, remain constant. We emphasize that the advantage of using
quasicrystals in this case is not in their arbitrarily-high symmetry,
but rather in the fact that there is no restriction on the
combinations of wave vectors that may appear in their reciprocal
lattices (provided that the symmetry of the quasicrystal is not of
particular importance~\cite{physa,RLlattices}).

The novel optical devices that we have been developing are based on
materials that facilitate the nonlinear interaction between light
waves in the form of three-wave mixing. These are processes in which
two incoming waves of frequencies $\omega_1$ and $\omega_2$ interact
through the quadratic dielectric tensor $\chi^{(2)}$ of the material
to produce a third wave of frequency $\omega_3=\omega_1\pm\omega_2$;
or the opposite processes in which a single wave spontaneously breaks
up into two.  Three-wave mixing is severely constrained in dispersive
materials, where $\omega(\kv)$ is not a linear function, because the
interacting photons must also conserve their total momentum. Even the
slightest wave-vector mismatch $\Delta\kv = \kv_1 \pm \kv_2 - \kv_3$
appears as an oscillating phase that averages out the outgoing wave,
giving rise to the so-called ``phase-matching problem.'' We have
explained how one could fully solve the most general phase-matching
problem using well-known ideas from the theory of
quasicrystals~\cite{Lifshitz_PRL2005}. The solution is based on the
idea that in crystals, whether periodic or not, continuous translation
symmetry is broken, as described above in Sec.~\ref{QCs}. As a
consequence, momentum conservation is replaced by the less-restrictive
conservation law of crystal momentum whereby momentum need only be
conserved to within a wave vector from the reciprocal lattice of the
crystal. The fabrication of an efficient frequency-conversion device
is therefore a matter of {\it reciprocal-lattice
  engineering}---designing an artificial crystal, from the quadratic
dielectric field of the material $\chi^{(2)}({\bf r})$, whose
reciprocal lattice contains any desired set of mismatch wave vectors
$\Delta\kv^{(j)}$, $j=1\ldots N$, required for phase matching any
arbitrary combination of $N$ three-wave mixing processes.

The idea of using a one-dimensional periodic modulation of the
relevant component of the quadratic dielectric tensor, for the purpose
of phase matching a single three-wave process, was suggested already
in the early 1960's~\cite{Armstrong1962,%
  Freund_NonlinearDiffraction1968,Bloembergen1970}, and is termed
``quasi-phase matching''. Since then this approach has been
generalized using more elaborate one-dimensional~\cite{Fejer_QPM1992,%
  Ming_THG1997,Keren_PRL2001} and
two-dimensional~\cite{Berger_NPC1998,Broderick_Hexagonal2000,%
  Bratfalean_NPQC2005,Ma_OcatgonalNPQC_APL2005} designs, but only as
{\it ad hoc\/} solutions for multiple processes.  We have demonstrated
that engineering the reciprocal lattice of a nonlinear photonic
quasicrystal to contain any desired set of mismatch vectors---a task
that 25 years of research in quasicrystals have taught us how to
solve---provides the most general solution for the long-standing
problem of multiple phase-matching~\cite{Lifshitz_PRL2005}. We
described elsewhere~\cite{alon1,alon2} a number of novel optical
devices that have actually been fabricated using these ideas, and
tested experimentally. These devices attest to the general nature of
the quasicrystal-based solution to the multiple phase-matching
problem.

\section{Towards self-assembly of quasicrystalline nanostructures --
  The recent discovery of soft-matter quasicrystals}
\label{soft}

An important development accured recently with the experimental
discovery that even soft matter can self-assemble into structures with
quasiperiodic long-range order.\footnote[2]{For the sake of historical
  accuracy, it should be noted that at some point the blue phase III
  of liquid crystals, also known as the ``blue fog'', was thought to
  have icosahedral quasicrystalline order~\cite{hornreich,rokhsar},
  but this eventually turned out not to be the
  case~\cite{wright,lubensky1}. Also, incommensurate helical
  twist-grain-boundary phases are known to exist in smectic liquid
  crystals~\cite{renn,goodby}, but the quasiperiodic order in this
  case is essentially only along the 1-dimensional screw axis.} In one
case, dendrimers that assume a conical shape assemble into micelles,
which then pack to form a perfect dodecagonal (12-fold)
quasicrystal~\cite{zeng1,zeng2,mehl}. In another case, $ABC$
star-shaped block terpolymers---in which the length ratios of the
three arms, $B/A$ and $C/A$, can be chemically-controlled---assemble
into a host of 2-dimensional columnar structures, one of which is,
again, a dodecagonal quasicrystal~\cite{takano1,matsushita,takano2}.
This phase has also been reproduced numerically using lattice Monte
Carlo simulations~\cite{doteraICQ9}. A similar square-triangle tiling
has also been observed in a liquid crystal composed of T-shaped
molecules~\cite{chen}, which forms yet a third soft system which may
potentially self-assemble into a dodecagonal quasicrystal. The
characteristic length of the basic building blocks ranges in these
systems from about $10$ to about $100$ nanometers---2 to 3 orders of
magnitude greater than the atomic length scales found in solid-state
quasicrystals. This property of soft quasicrystals is what will
potentially make them useful as functional self-assembled
nanomaterials~\cite{scaffolds,electronics,selfassemblyreview}, and at
the same time as a new experimental platform for detailed---real-space
and real-time---study of quasiperiodic long-range order.

Investigations of these new soft members of the quasicrystal family of
materials, are only at their infancy. For example, even the space
groups of the observed phases have not been determined, although from
the diffraction patterns of the dendrimer liquid crystals given by
Zeng {\it et al.}~\cite{zeng1,zeng2} it seems that they have a 12-fold
screw axis, and therefore, most likely, the nonsymmorphic space group
$P12_6/mcm$~\cite{rabson}.  More generally, the same
questions~\cite{mayou-discussion,thiel-discussion,widom-discussion}
concerning the thermodynamic stability, the role of clusters in
formation and dynamics, and the importance of phasons, apply to soft
quasicrystals as they do to hard quasicrystals.  Yet the answers may
be more tractable (albeit possibly different as the systems are quite
different). Thus, the study of soft quasicrystals will clearly have
implications well beyond the limits of the specific soft systems that
have been discovered so far, and is likely to promote the fundamental
understanding of quasicrystals in general. Fortunately, the study of
soft quasicrystals is happening at a point in time when the science of
quasicrystals is ready and mature to tackle these newly discovered
systems. We are no longer taken by surprise whenever a new chemical or
physical system exhibits quasicrystalline structure. We are prepared
with the appropriate tools to study and explore it, and hopefully also
to exploit it for the control of the self-assembly of useful
nanomaterials.

\section{Insights from an even softer system -- Quasicrystalline
  surface waves}  
\label{LP}

Motivated by experiments with parametrically-excited surface waves
(Faraday waves), exhibiting dodecagonal quasiperiodic
order~\cite{edwards}, Lifshitz and Petrich~\cite{faraday} developed
some years ago a model for describing the pattern-forming dynamics of
a two-dimensional field in which two length scales undergo a
simultaneous instability.  This model is an extension of the
Swift-Hohenberg equation~\cite{swift}, which is used for describing a
variety of different pattern-forming systems~\cite{cross}.  Its
dynamics is relaxational, $\partial_t \rho = -\delta {\cal F} / \delta
\rho$, driving a 2-dimensional field $\rho(x,y,t)$ towards the minimum
of an ``effective free energy'' (\ref{eq:free}),
\begin{equation}
 \label{eq:lyapunov}
 {\cal F}_{LP}\{\rho\} = \int \!dx\, dy\, \bigl\{- \frac12 \varepsilon
 \rho^2 + \frac12 
 [(\nabla^2+1)(\nabla^2+q^2)\rho]^2 
 - \frac13 \alpha \rho^3 + \frac14 \rho^4 \bigr\},
\end{equation}
yielding a dynamical equation of the form
\begin{equation}
 \label{lpeqn}
 \partial_t \rho = \varepsilon \rho - (\nabla^2 + 1)^2(\nabla^2 +q^2)^2 \rho
 +\alpha \rho^2 - \rho^3.
\end{equation}
It essentially mimics the dynamics of a generic 2-dimensional material
in search of its ground state, and therefore offers us important
insight and a good starting point for our current investigation of
soft quasicrystals. A Java simulation of the dynamical
equation~(\ref{lpeqn}), starting from random initial conditions, and
arriving at a quasicrystalline pattern can be found at {\tt
  http://www.its.caltech.edu/$\sim$mcc/Patterns/Demo4\_6.html}.

The Lifshitz-Petrich free energy ${\cal F}_{LP}$ is indeed generic,
imposing only two requirements on a material, described by a
2-dimensional density $\rho(x,y,t)$: (a) The existence of two
characteristic length scales, whose ratio is given by the parameter
$q$; and (b) The existence of effective 3-body interactions, whose
importance is given by the relative strength of the parameter
$\alpha$. In~\cite{faraday} we were able to show analytically (using
standard methods~\cite[ch.~4.7]{Chaikin} and~\cite{gronlund}), and
demonstrate numerically, that if $q$ is chosen around $2\cos(\pi/12) =
\sqrt{2+\sqrt3} \simeq 1.932$ one can obtain a ground state with
quasiperiodic long-range order and dodecagonal symmetry, yet no choice
of $q$ yields globally-stable ground states with octagonal or
decagonal symmetry.  The latter two have insufficient triplets of wave
vectors in the Fourier Lattice $L$ [Eq.~\(eq:fourier)] that add up to
zero to overcome the cost of additional density modes, as compared
with the hexagonal state. Thus, in two dimensions, the requirements of
two length scales and 3-body interactions are sufficient to stabilize
dodecagonal quasicrystals, but insufficient to stabilize octagonal or
decagonal quasicrystals.  This raises the possibility that the fact
that the soft quasicrystals discovered to date are all dodecagonal,
may be accounted for using a free energy similar to ${\cal F}_{LP}$.
Note that for hard quasicrystals the situation is
different---decagonal quasicrystals are thermodynamically stable
whereas octagonal and dodecagonal quasicrystals are believed to be
metastable---indicating that the stabilization mechanism for soft
quasicrystals might be quite different from that of hard
quasicrystals.

\section{Validity of density-wave theories of quasicrystals}
\label{free}

At the outset, as we argue in more detail
elsewhere~\cite{lifshitz-diamant}, the experimental soft systems in
which quasicrystalline order has been observed seem to satisfy the
basic assumptions of the Lifshitz-Petrich theory described in
Sec.~\ref{LP}. The asymmetric and heterogeneous structure of the star
polymers and dendrimers will most likely require more than one length
scale for an appropriate coarse-grained
description.\footnote[3]{Indeed, coarse-grained free energies
  previously used for amphiphilic self-assembly~\cite{Gompper} involve
  more than one characteristic length scale due to the asymmetry of
  the molecules and the resulting tendency to form curved interfaces.}
Their ultra-soft repulsion and resulting strong inter-penetration
\cite{Lowen1,Lowen2,Lowen3,Lowen4,Lowen7,Lowen9,Kamien3} imply that
3-body interactions should be significant \cite{Lowen6}.  Thus, we
expect that studies that we are currently undertaking will yield
functionals similar in nature to ${\cal F}_{\rm LP}$ of
Eq.~(\ref{eq:lyapunov}).  Significant differences may emerge,
nonetheless, as the systems considered here are 3-dimensional and
differ in their microscopic structure.  For instance, two order
parameters rather than one might be required \cite{takano2}, which
could potentially allow point-group symmetries other than dodecagonal
to be observed.\footnote[4]{Models with two order parameters were
  suggested also for hard quasicrystals~\cite{Mermin85} and
  pattern-forming systems~\cite{Muller}, yielding additional
  quasicrystalline ground-state symmetries.}

Another key insight can be drawn from a recent theoretical
observation, according to which dispersions of soft, fuzzy, particles
are essentially different in their thermodynamics from those of hard
particles \cite{Kamien1,Kamien2}. The overlap of the soft ``coronas''
surrounding the particles leads to a driving force acting to minimize
their interfacial area, in analogy with foams.  Consequently, unusual
liquid-crystalline structures can be stabilized in systems of soft
spheres \cite{Kamien1,Kamien2,li,Kamien4}. Both star polymers and
flexible dendrimers fall into this fuzzy category~\cite{dotera,cho},
yet they may be highly aspherical. Likos {\it et al.}~\cite{Lowen9}
have also shown that stars and flexible dendrimers have the same kind
of soft pair potentials.  We thus expect such considerations of
interfacial-area minimization to become highly relevant in the
study of soft quasicrystals.

A 3-dimensional version of an LP-like free energy may remind the
reader of the early attempts by Kalugin, Kitaev, and
Levitov~\cite[KKL]{kalugin}, who extended the model of Alexander and
McTague~\cite{alexander}, using density-wave theories to establish
that the icosahedral quasicrystal has lower free energy than the
competing {\it bcc\/} phase.  Narasimhan and Ho~\cite[NH]{narasimhan}
managed to show in their model that there are regions in parameter
space in which a dodecagonal quasicrystal is favored and other regions
in which a decagonal quasicrystal is favored. These attempts were
eventually discontinued after it was shown by Gronlund and
Mermin~\cite{gronlund} that the addition of a quartic term to the
cubic free energy of KKL reverses the outcome of the calculation,
establishing the {\it bcc\/} phase as the favored one. For hard
crystals it is unclear where to truncate the density-wave expansion of
the free energy and whether such a truncation is fully justified. As
we discussed in~\cite{lifshitz-diamant}, for the soft systems considered
here the truncation of the expansion should be more valid.  We are
therefore in a position now to reexamine some of the old conclusions,
based on density-wave theories of quasicrystals, as they are likely to
apply to \emph{soft} quasicrystals.  Roan and Shakhnovich~\cite{roan}
performed such a stability study for the case of icosahedral order in
diblock copolymers and concluded that such order is only metastable.
Nevertheless, we are encouraged by the old results of NH who
established the stability of dodecagonal, as well as decagonal,
quasicrystals within the same model.

\section{Dislocation and phason dynamics -- From soft to photonic
  quasicrystals} 
\label{dynamics}

Valuable knowledge about the nature of quasiperiodic order, important
also for the control of its self-assembly, can be obtained by studying
its topological defects~\cite{anderson,sethna,mermindefects}, and its
low-energy collective excitations---in particular those associated
with the phason degrees of freedom. Much like phonons, phasons are
low-energy excitations of the quasicrystal, only that instead of
slightly shifting the atoms away from their equilibrium positions, the
relative positions of atoms are interchanged. Their existence stems
directly from the fact that the dimension $D$ of the order parameter
space $V^*/L^*$ is greater than the physical dimension $d$. Thus, in
addition to $d$ independent (acoustic) phonon modes there are $D-d$
independent phason modes.

The existence of phasons as fundamental degrees of freedom, affecting
the physical behavior of quasicrystals, has been clearly established
over the years. Their role in a dynamical density-wave theory of
quasicrystals was developed in a series of
papers~\cite{ph1,ph2,ph3,ph4,ph5,ph6,ph8,bancel} immediately following
the announcement of the discovery of quasicrystals. Phasons have been
observed in numerous experiments, whether directly or indirectly,
throughout the past two decades~\cite[for
example]{kuo1,kuo2,marc,edagawa,sonia}, and are still a source of
interesting analytical puzzles~\cite{ted} and ongoing
debate~\cite{widom-discussion}.

We have recently begun investigating the motion of dislocations and
the dynamics of phasons in the dodecagonal ground state of the LP
equation~\cite{barak2}. We are studying, both analytically and
numerically, such questions as the climb velocity of dislocations
under strain, the pinning of dislocations by the underlying
quasiperiodic structure under conditions of weak diffusion, and the
relaxation of phason strain as two dislocations of opposite
topological sign merge and annihilate each other. These studies are
impossible to conduct with either Faraday waves or hard atomic
quasicrstals. Thus soft quasicrystals, with their $10nm$ to $100nm$
length scales, may become one of the first natural experimental system
to provide real quantitative microscopic answers regarding the
dynamics of the fundamental degrees of freedom in a
quasicrystal---defects and low-energy excitations.

In the meantime we have embarked on the study of these unique
dynamical degrees of freedom in a facsinating new artificial form of
quasicrystalline medium---an optically-induced nonlinear photonic
quasicrystal---which we have recently demonstrated~\cite{nature}.  In
this systems beams of light inetract nonlinearly by changing the index
of refraction of a photorefractive material. Their dynamics is
governed by a different type of equation---the so-called nonlinear
Schr\"{o}dinger equation. Nevertheless, it is capable of stabilizing
structures with quasicrystalline order where the typical distance
between crystal sites is $15\mu m$ to $30\mu m$. This has allowed us
to study the microscopic dynamics of dislocations~\cite{nature} as
well as phasons~\cite{natmat}. These artificial systems already
provide useful information reagrding the dynamics of fundamental
degrees of freedom in quasicrystals.  Similar investigations of soft
quasicrystals, should provide valuable insight into their physical
nature, as well as that of all quasicrystals, regardless of the
physical or chemical system in which they are realized. This insight
will be most valuable in trying to design and control the self
assembly of quasiperiodic nanomaterials.

\section*{Acknowledgments} I would like to thank Alexey Madison, Yuri
Magarshak, and Sergey Kozirev for inviting me to St.~Petersburg, where
I was encouraged to stress the connections between my different fields
of research---nanotechnology and quasicrystals. I would like to thank
Haim Diamant and Kobi Kan with whom I am currently trying to
understand the stability of soft quasicrystals and how one could
control their self assembly. I thank Gilad Barak with whom I have
studied the dynamics of dislocations and phasons in the
Lifshitz-Petrich equation. I thank Ady Arie, Alon Bahanad, and Noa
Voloch with whom I have been developing the use of nonlinear photonic
quasicrystals for the purpose of frequency conversion.  Finally, I
would like to thank Barak Freedman and Mordechai Segev with whom I
have been studying the dynamics of optically-induced nonlinear
photonic quasicrystals. This work is funded by the Israel Science
Foundation under Grant No.~684/06.

\end{document}